\documentclass[9.5pt,journal,final,finalsubmission,twoside]{IEEEtran}

\usepackage[english]{babel}
\usepackage{amsmath,amssymb}
\usepackage{times}
\usepackage{relsize}
\usepackage{graphicx}
\usepackage{url}
\usepackage{booktabs}
\usepackage{multirow}
\usepackage{mparhack}
\usepackage{subfigure}
\usepackage{authblk}
\usepackage{amsthm}

 \usepackage{algorithm}
\usepackage{algorithmicx}
 \usepackage{algpseudocode}

\usepackage{epstopdf}

\usepackage{array}
\usepackage{cite}
\usepackage{eqparbox}
\usepackage{mdwmath}
\usepackage{balance}
\usepackage{epsfig}
\usepackage{xcolor}

\usepackage{mathtools}
\usepackage{bm}
\usepackage{amsfonts}
\usepackage[all]{xy}
\usepackage{etoolbox}
\usepackage{graphicx}
\DeclareMathAlphabet{\mathantt}{OT1}{antt}{li}{it}
\DeclareMathAlphabet{\mathpzc}{OT1}{pzc}{m}{it}

\usepackage{accents}
\DeclareFontFamily{OT1}{pzc}{}
\DeclareFontShape{OT1}{pzc}{m}{it}%
  {<-> s * [1.1] pzcmi7t}{}
\DeclareMathAlphabet{\mathpzc}{OT1}{pzc}%
                     {m}{it}

\makeatletter
\patchcmd{\@maketitle}
  {\addvspace{0.5\baselineskip}\egroup}
  {\addvspace{-1.45\baselineskip}\egroup}
  {}
  {}
\makeatother

\title{Data-aided Active User Detection with False Alarm Correction in Grant-Free Transmission}

\author{Linjie~Yang,~\IEEEmembership{Student Member,~IEEE}, Pingzhi~Fan,~\IEEEmembership{Fellow,~IEEE},  Des~McLernon,~\IEEEmembership{Member,~IEEE}, 
Li X~Zhang,~\IEEEmembership{}
	\thanks{Linjie Yang and Pingzhi Fan are with the School of Information Science and Technology, Southwest Jiaotong University, Chengdu 610031, China (e-mail: yanglinjie@my.swjtu.edu.cn; pzfan@swjtu.edu.cn).}
	\thanks{
	Des McLernon and Li X. Zhang  are with the School of Electronic and Electrical Engineering, University of Leeds, Leeds LS2 9JT, UK	(e-mail: D.C.McLernon@leeds.ac.uk, L.X.Zhang@leeds.ac.uk )}
}

\begin{document}

\maketitle

\begin{abstract}
In most existing grant-free (GF) studies, the two key tasks, namely active user detection (AUD) and payload data decoding, are handled separately. 
In this paper, a two-step data-aided AUD scheme is proposed, namely the initial AUD step and the false alarm correction step respectively.
To implement the initial AUD step, an embedded low-density-signature (LDS) based preamble pool is constructed.
In addition, two message passing algorithm (MPA) based initial estimators are developed.
In the false alarm correction step, a redundant factor graph is constructed based on the initial active user set, on which MPA is employed for data decoding.
The remaining false detected inactive users will be further recognized by the false alarm corrector with the aid of decoded data symbols.
Simulation results reveal that both the data decoding performance and the AUD performance are significantly enhanced by more than $1.5$ dB at the target accuracy of $10^{-3}$ compared with the traditional compressed sensing (CS) based counterparts.

\end{abstract}
\begin{IEEEkeywords}
 Grant free, False alarm correction, MPA
\end{IEEEkeywords}

\IEEEpeerreviewmaketitle

\section{Introduction}
Massive machine-type communication (mMTC) is one of the most
popular services in fifth-generation (5G) mobile communication systems.
Since conventional orthogonal multiple access (OMA) cannot meet the explosive demand due to the limited orthogonal resources, non-orthogonal multiple access (NOMA) technologies are advocated  to support the massive connectivity.
Among the many available NOMA schemes, low-density-signature orthogonal frequency division multiplexing (LDS-OFDM) \cite{shahab2020grant} is one of the most generic solutions in code domain.
In an LDS-OFDM system, the  message passing algorithm (MPA) with near-optimal performance is employed to cancel
interference among multiple users. 
Benefiting from the LDS structure, the complexity of the MPA algorithm becomes affordable.
However, the MPA algorithm is implemented on the underlying assumption that each user's activity information is perfectly known at the base station (BS).
However, in massive IoT networks, this assumption is impractical.

Now in 5G New Radio, the approval proposed to reduce latency is grant-free(GF) random access. This means channel resources can be accessed without being arranged through a handshaking process. 
To realize the GF requirement of LDS-OFDM system, there are two mainstream solutions widely studied. Firstly, a framework referred to as compressed sensing based MPA (CS-MPA) detector is proposed where active users will transmit their specific non-orthogonal
preamble with length $L_{p}$ before their data transmission begins \cite{oyerinde2019compressive,wang2016dynamic}. By leveraging users' activity sparsity, the active user detection (AUD) task is formulated as a standard CS problem and solved by the existing CS recovery algorithms efficiently, e.g. orthogonal matching pursuit (OMP) \cite{oyerinde2019compressive}, dynamic compressed sensing (DCS)\cite{wang2016dynamic}, and approximate message passing algorithm (AMP) \cite{yang2021cross,8264818} etc. Then, MPA is performed to reliably detect the transmitted symbols of the active users.
On the other hand, some researchers propose to add an extended zero constellation point into the conventional LDS constellation alphabet \cite{zhu2010exploiting}. The key idea is that one can recognize the activity states of users through their decoded symbols, i.e.
if the detected zero symbols in a user's packet is large enough, this user is considered as inactive. 

However,  
in a CS-MPA detector, the AUD and data payload decoding are normally handled separately. 
The feasibility that error correction with the aid of decoded data symbols provides additional mechanism for performance improvement is ignored \cite{bian2021supporting}.
In \cite{zhu2010exploiting},  
MPA is directly employed to decode data symbols of all potential users, at the absence of activity state information of potential users in the cell.
But, the complexity of this approach would become prohibitive upon the increase of the potential user number.

Based on the above discussion, 
Our main contributions in this paper are summarized as follows.
\begin{itemize}
	\item Firstly, 
	a data-aided two-step AUD scheme is proposed. In step 1,
	 an initial active user set which 
contains a small number of false alarms is estimated by the initial estimator from the received preamble signal. 
In step 2, these false alarms are further corrected by the designed false alarm corrector. 
		
	\item Secondly, to estimate an initial active user set,
an embedded LDS based preamble pool is firstly constructed. 
Then, an MPA based initial estimator is presented. To reduce the complexity of the MPA detector, a traffic load aided MPA (TL-MPA) based detector is further proposed.
	
	\item Finally, 
	based on the fact that if a user is inactive, the number of detected zero symbols should be large,
	a false alarm corrector based on multiple zero symbol detection is implemented in the data decoding process to peel off the remaining false alarms in the initial active user set.

\end{itemize}
The rest of the paper is orgnised as follows.
System model is introduced in Section II. 
The construction method of embedded LDS based user preamble and two MPA based initial estimators are depicted in Section III.
 In Section IV, the proposed false alarm corrector is described. 
 Complexity analysis is provided in Section V.
 Simulation results are presented in Section VI. 
Finally, the paper is concluded in Section VII.

\section{System model}
\begin{figure}[htbp]
	\centering
	\includegraphics[scale=0.7]{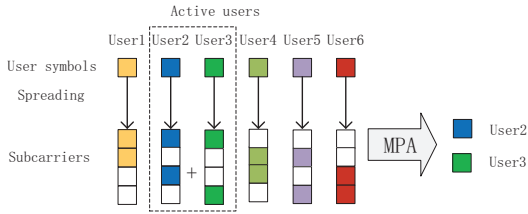}
	\caption{Graphic representation of the GF LDS-OFDM, where potential user number $N=6$, active user number $N_{a}=2$, sub-carriers number $L_{s}=4$.}
	\label{1}
\end{figure}

An up-link scenario of a single IoT cell with $N$ IoT devices
each of which
is assigned a user-specific preamble, is considered. Only $N_{a}$ users are active at any given time point. The sparsity $\lambda$ is defined as $\lambda = \frac{N_{a}}{N}$. 
Meanwhile, only a single antenna is considered at both the user side and base station (BS) for low-cost IoT.
Moreover, for the simplicity, 
we consider AWGN channel and assume perfect symbol-wise synchronization.

%% 2021-11-26
In this paper, the whole transmission period contains two stages, namely the preamble transmission stage and the data transmission stage.
In the preamble transmission stage,
for user $u, 1 \leq u \leq N$, once it becomes active, its activity state changes to $a_{u}=1$ from  $a_{u}=0$. Then, its user-specific preamble $\mathbf{s}_{u}$ with length $L_{p}$ is transmitted to the BS in a GF manner\cite{yang2021cross,8264818}. 
The construction of a users' preamble pool will be elaborated later in Section III.
After that, its data transmission stage begins.
Firstly, its data packet $\mathbf{x}_{u}=[x_{u}[1],x_{u}[2],\cdots,x_{u}[K]]\in \mathbb{C}^{K\times 1}$ is prepared. The $k^{th}$ data symbol $x_{u}[k],1\leq k \leq K$ is selected from the alphabet $\mathcal{X}_{u} = \mathcal{X}\cup\{0\}$. 
The original alphabet $\mathcal{X}$ is generated according to \cite{van2009multiple} where each
standard $M-$ary phase shift keying (PSK) symbol is multiplied by a user-specific complex coefficient.
The cardinality of the final alphabet is $M+1$. 
Then, $\mathbf{x}_{u}$ is modulated onto a user specific signature sequence $\mathbf{c}_{u}$ with length $L_{s}$. 
All active users' signals are superposed and propagated simultaneously over $L_{s}$ sub-carriers.
In this paper, the signature sequence $\mathbf{c}_{u}, 1\leq u \leq N$ is selected from the columns of the parity check matrix of a regular low-density-parity-check (LDPC) code.
For the example in Figure. \ref{1}, the LDS signature matrix $\mathbf{C}_{4,6}=[\mathbf{c}_{1},\mathbf{c}_{2},\cdots, \mathbf{c}_{6}]$ is given by
\begin{equation}
\mathbf{C}_{4,6} = \left[
\begin{array}{cccccc}  
1 &1 &1  &0         &0 &0 \\
1 &0     &0      &1      &1     &0\\
0 &1     &0      &1      &0        &1\\
0 &0         &1  &0        &1 &1
\end{array} 
\right].
\end{equation}

At the receiver side, the received signal of users' preamble can be modeled as
\begin{equation}
\mathbf{y}_{p} = \sum_{u=1}^{N} a_{n}\mathbf{s}_{n} + \mathbf{n}_{p}.
\end{equation}
The received signal of $k^{th}, 1\leq k \leq K$ data symbol of user $u$ on the $l_{s}^{th}$ sub-carrier is modeled as
\begin{equation}
\mathbf{y}[l_{s},k] = \sum_{u=1}^{N}a_{u}\mathbf{c}_{u}[l_{s}]\mathbf{x}_{u}[k] + \mathbf{n}_{d}[l_{s}, k],
\end{equation}
where $\mathbf{c}_{u}[l_{s}], 1\leq l_{s}\leq L_{s}$ denotes the $l_{s}^{th}$ component of $\mathbf{c}_{u}$.
$\mathbf{n}_{p} \in \mathbb{C}^{L_{s}\times 1}$ and $\mathbf{n}_{d} \in \mathbb{C}^{L_{s}\times K}$ represent background noise which obey i.i.d. Gaussian distribution $\mathcal{CN}(0,\sigma^{2})$. 
Additionally, as stated in \cite{wang2016dynamic,oyerinde2019compressive}, the preamble length equals the sub-carrier number, i.e. $L_{s}=L_{p}$ in this paper.

\section{Initial Active Uset Set Detection}

\subsection{ Embedded LDS based User Preamble Construction}

To obtain the initial active user set, which is also referred to as the super-set $\hat{\mathcal{U}}_{ac}^{I}$ (i.e. $\mathcal{U}_{ac}\subset \hat{\mathcal{U}}_{ac}^{I}$ \cite{mazumdar2021support}), only a test on each sub-carrier is required \cite{inan2019group}. 
If there is no user transmitting data on the $l_{s}^{th}$ sub-carrier, the $l_{s}$ sub-carrier is idle and the outcome of the test is $\mathbf{Y}_{t}[l_{s}]=0$. Otherwise, the $l_{s}$ sub-carrier is busy and $\mathbf{Y}_{t}[l_{s}]=1$. Then, based on $\mathbf{Y}_{t}$, $\hat{\mathcal{U}}_{ac}^{I}$ can be efficiently estimated with a cover decoder by directly removing the inactive users, i.e. $\{u \vert \mathbf{C}[l_{s},u]=1, Y_{t}[l_{s}]=0\}$ \cite{inan2019group}.
%For the example in Fig. \ref{1}, the signature sequences of user 2 and user 3 are $\mathbf{c}_{2}=[1,0,1,0]^{T},\mathbf{c}_{3}=[1,0,0,1]^{T}$. $(\cdot)^{\rm T}$ denotes the transpose operation. Then, the perfect estimation of sub-carrier tests $\mathbf{Y}_{t}$, $\hat{\mathbf{Y}}_{t}=[1,0,1,1]^{T}$. The second sub-carrier is idle.
%It implies that user 1, user 4 and user 5 are surely inactive, i.e. $\hat{\mathcal{U}}_{ina}=\{u \vert \mathbf{C}[2,u]=1\}$. As a result,  $\hat{\mathcal{U}}_{ac}^{I}=\{2,3,6\}$ is obtained as the complementary set of  $\hat{\mathcal{U}}_{ina}$.

Informed by \cite{inan2019group}, in our case, 
the target of the user preamble $\mathbf{s}_{u}, 1\leq u\leq N$ is converted to convey 
the non-zero elements' positions in the signature sequence $\mathbf{c}_{u}$, rather than conveying its identity directly.
To this end, we first generate a Zad-off Chu (ZC) sequence $\mathbf{z}_{r}[n]$ and its $L_{s}-1$ cyclic-shifting versions $\mathbf{z}_{r}[n+1],\cdots,\mathbf{z}_{r}[n+L_{s}-1]$ to form a ZC sequence set, $\{\mathbf{z}_{r}[n],\mathbf{z}_{r}[n+1]\cdots\mathbf{z}_{r}[n+L_{s}-1]\}$.
$\mathbf{z}_{r}[n]= \exp[-j\pi r n(n+1)/L_{s}], n=0,1,\cdots, L_{s}-1$ denote a ZC sequence with root number $r$ \cite{jang2016early}.
Then, for user $u, 1\leq u \leq N$, each $l_{s}^{th}$ ZC sequence is selected once $\mathbf{c}_{u}[l_{s}]=1$. Finally, combine these selected ZC sequences to form the preamble of user $u$, $\mathbf{s}_{u}$.
For example, for user 1 whose signature sequence is $\mathbf{c}_{1}=[1,1,0,0]^{T}$, its preamble is constructed as $\mathbf{s}_{1}=\frac{1}{\sqrt 2}( \mathbf{z}[n] +  \mathbf{z}[n+1])$.
In the same spirit, the embedded LDS based preamble construction is summarized as 
\begin{equation}
\mathbf{s}_{u} = \frac{1}{\sqrt{w_{c}}}\sum_{l_{s}=1}^{L_{s}} \mathbf{c}_{u}[l_{s}] \mathbf{z}_{r}[n+l_{s}-1].
\end{equation}
Lastly, the column weight $w_{c}$ of $\mathbf{C}$ is introduced to normalize the unit power of the constructed user preamble.

\subsection{MPA based initial estimator}
At the receiver, the correlation value between the preamble received signal $\mathbf{y}_{p}$ and the aforementioned $L_{s}$ reference ZC sequences $\mathbf{z}_{r}[n+l_{s}], 0\leq l_{s} \leq L_{s}-1$ in Section III is first calculated.
\begin{equation}
\mathbf{R}[l_{s}] = \frac{\sqrt{w_{c}}}{{N_{zc}}}\vert \sum_{n=0}^{N_{zc}-1}\mathbf{y}_{p}\mathbf{z}_{r}^{\star}[n+l_{s}]\vert,
\end{equation}
where $(\cdot)^{\star}$ is the complex conjugate operator.
Then, $\mathbf{Y}_{t}$ can be estimated as 
\begin{equation}
\hat{\mathbf{Y}}_{t}[l_{s}] =  \left\{\begin{array}{l}
1,\quad  \mathbf{R}[l_{s}]\geq\tau_{zc}\\
0, \quad {\rm otherwise}
\end{array}\right.
\end{equation}
where $\tau_{zc}$ is a predefined threshold. 
Based on $\hat{\mathbf{Y}}_{t}$, $\hat{\mathcal{U}}_{ac}^{I}$ can be  estimated by a cover decoder \cite{inan2019group}. 

However, a cover decoder will only makes a hard decision according to $\mathbf{Y}_{t}$, which is sensitive to the background noise.
To improve the robustness of the initial estimator, an MPA based detector is proposed.
Define the traffic load  of the $l_{s}^{th}$ sub-carrier as $\mathbf{Y}_{l}[l_{s}]$. $\mathbf{Y}_{l}[l_{s}]= w$ indicates there are exactly $w$ users which are transmitting data on the $l_{s}^{th}$ sub-carrier.    
According to \cite{jang2016early}, the values of $\mathbf{R}$ obey a Rice distribution, i.e.  $\mathbf{R}[l_{s}]\sim {\rm Rice}(\mathbf{Y}_{l}[l_{s}] ,\frac{\sigma}{\sqrt{2N_{zc}} )  })$. We denotes the probability density function of the Rice distribution as
\begin{equation}
{\rm Rice}(x\vert A,\sigma)= \frac{x}{\sigma^{2}} \exp(-\frac{x^{2}+A^{2}}{2\sigma^{2}})I_{0}(\frac{xA}{\sigma^{2}}),
\end{equation} 
where $I_{0}(\cdot)$ is the modified Bessel function of the first kind with the order zero.
The rules of the proposed MPA detector in the $i^{th}$ iteration are given by 
\begin{equation}
\begin{split}
&E_{l_{s} \rightarrow u}^{(i)}( a_{u} ) =  \mathcal{R} \sum\limits_{ \mathclap{{u' \in \mathcal{N}(l_{s})\backslash u} \atop a_{u'}\in \{0,1\}}  } {\rm Rice}( \mathbf{R}[l_{s}] \ \vert \ a_{u}+\sum_{ \mathclap{	u'\in \mathcal{N}(l_{s})\backslash u}
}a_{u'}, \\ 
&{\sigma}/{\sqrt{2N_{zc}}} ) 
\cdot\prod_{ \mathclap{u'\in \mathcal{N}(l_{s})\backslash u}  }E_{u'\rightarrow l_{s}}^{(i-1)}(a_{u'}),
\end{split}
\end{equation}
\begin{equation}
E^{(i)}_{u\rightarrow l_{s}}(a_{u})=\prod_{\mathclap{l_{s}'\in \mathcal{N}(u)\backslash l_{s}}}E_{l_{s}'\rightarrow u}^{(i-1)}(a_{u}),
\end{equation}
where $E_{l_{s} \rightarrow u}^{(i)}( a_{u} )$ denotes the extrinsic information passed from the $l_{s}^{th}$ check node to the $u^{th}$ variable node in the $i^{th}$ iteration.
$E^{(i)}_{u\rightarrow l_{s}}(a_{u})$ denotes the extrinsic information passed from the $u^{th}$ variable node to the $l_{s}^{th}$ check node in the $i^{th}$ iteration.
Then constant $\mathcal{R}$ is chosen such that $E_{l_{s} \rightarrow u}^{(i)}( a_{u}=0 )+E_{l_{s} \rightarrow u}^{(i)}( a_{u}=1 )=1$. $\mathcal{N}(u)$ denotes the set of sub-carriers  occupied by user $u$. $\mathcal{N}(u)\backslash l_{s}$ means excluding the $l_{s}^{th}$ sub-carrier from $\mathcal{N}(u)$. Similarly,  $\mathcal{N}(l_{s})$ denotes the set of users occupying the $l_{s}^{th}$ sub-carrier. $\mathcal{N}(l_{s})\backslash u$ denotes excluding user $u$ from  $\mathcal{N}(l_{s})$. The MPA based detector is initialized by $E^{(0)}_{u\rightarrow l_{s}}(a_{u}=1)=\lambda$ and $E^{(0)}_{u\rightarrow l_{s}}(a_{u}=0)=1-\lambda$.
Then, the \emph{a posterior} probability whether user $u$ is active is computed as
\begin{equation}
E(a_{u})=\prod_{\mathclap{l_{s}'\in \mathcal{N}(u) }}E_{l_{s}'\rightarrow u}^{(i)}(a_{u}),
\end{equation}
Note that in our scheme, the missing detection should be avoided as much as possible \cite{inan2019group}. Hence, the decision rule of the proposed MPA based detector is 
\begin{equation}
\hat{a}_{u}=\left\{\begin{array}{l}
0,\quad  E(a_{u}=0)>0.99\\
1, \quad {\rm otherwise}
\end{array}\right.
\end{equation}

\subsection{Traffic load aided MPA (TL-MPA) based initial estimator}

The search space of the proposed MPA based detector in (8) is in the order of $\mathcal{O}(2^{w_{r}})$. $w_{r}$ denotes the row weight of $\mathbf{C}$.
The search space can be further reduced.
Now, 
the traffic load of $l_{s}^{th}$ sub-carrier $\mathbf{Y}_{l}[l_{s}]$ is estimated as
\begin{equation}
\hat{\mathbf{Y}}_{l}[l_{s}] = \left\{ \begin{array}{l}
\lfloor \mathbf{R}[l_{s}] \rfloor, \quad \ \ \mathbf{R}[l_{s}]-\lfloor \mathbf{R}[l_{s}] \rfloor < \tau_{zc} \\
\lfloor \mathbf{R}[l_{s}] \rfloor+1,  \mathbf{R}[l_{s}]-\lfloor \mathbf{R}[l_{s}] \rfloor \geq \tau_{zc}
\end{array}
\right.
\end{equation}
where $\lfloor \cdot \rfloor$ denotes the round down to the nearest integer. 
In the detection process, we only search the possible combinations such that  $a_{u}+\sum_{ {	u'\in \mathcal{N}(l_{s})\backslash u}
}a_{u'} = \hat{\mathbf{Y}}_{l}[l_{s}]$ on the $l_{s}^{th}$ sub-carrier.
The search space is reduced to the order of $\mathcal{O}( \binom{\hat{\mathbf{Y}}_{l}[l_{s}]}{w_{r}}  ) $ where $\binom{k}{n}$ denotes the number of combinations of $n$ items taken $k$ at a time.
Accordingly, the decoding rules of TL-MPA are given by
\begin{equation}
E_{l_{s}\rightarrow u}^{(i)} = \log(  \sum_{{\sum a_{u'}=\hat{\mathbf{Y}}_{l}[l_{s}]-1 }  }  \prod_{u'\in \mathcal{N}(l_{s})\backslash u   }p^{a_{u'}}(1-p)^{1-a_{u'}}   ),
\end{equation}
\begin{equation}
E_{u\rightarrow l_{s}}^{(i)} = \sum_{\mathclap{l_{s}' \in \mathcal{N}(u)\backslash l_{s}}} E_{l_{s}' \rightarrow u}^{(i-1)},
\end{equation}
where $p =   \frac{    \exp(E_{u'\rightarrow l_{s}}^{(i-1)})       }{ 1+\exp(E_{u'\rightarrow l_{s}}^{(i-1)})  } $. Particularly, $E_{l_{s}\rightarrow u}^{(i)}=-\infty$ if $\hat{\mathbf{Y}}_{l}[l_{s}]=0$.
The log-likelihood ratio (LLR) $\log(\frac{{\rm P}(a_{u}=1)}{{\rm P}(a_{u}=0)})$ is computed as 
\begin{equation}
r_{u} = \sum_{\mathclap{l_{s}' \in \mathcal{N}(u)}} E_{l_{s}' \rightarrow u}^{(i)},
\end{equation}
and $E_{u\rightarrow l_{s}}^{(0)}$ is initialized as $\log (\frac{\lambda}{1-\lambda})$.
Similar to the MPA based detector, the decision rule of TL-MPA is 
\begin{equation}
\hat{a}_{u}=\left\{\begin{array}{l}
0,\quad  r_{u}<-10,\\
1, \quad {\rm otherwise}
\end{array}\right.
\end{equation}

\section{Data-aided False Alarm Corrector}

After obtaining the initial active user set $\hat{\mathcal{U}}_{ac}^{I}$, a redundant factor graph $\mathcal{G}( \hat{\mathcal{U}}_{ac}^{I} )$ can be constructed by regarding the sub-carriers as the check nodes and users in $\hat{\mathcal{U}}_{ac}^{I}$ as the variable nodes. Then, the MPA algorithm \cite{hoshyar2008novel} can be employed to perform data decoding over the factor graph  $\mathcal{G}( \hat{\mathcal{U}}_{ac}^{I} )$\cite{zhu2010exploiting}. The decoding process of the $k^{th}, 1\leq k \leq K$ data symbol is formulated as
\begin{equation}
\hat{\mathbf{x}}_{u}[k] = {\rm MPA}( \mathbf{y}[:,k], \mathcal{G}(\hat{\mathcal{U}}_{ac}^{I}) ), u\in \hat{\mathcal{U}}_{ac}^{I}.
\end{equation}
However, the existence of the redundant variable nodes in $\mathcal{G}( \hat{\mathcal{U}}_{ac}^{I} )$ would degrade the decoding performance of MPA.
Hence, removing these redundant variable nodes is of great importance, and this motivates our false alarm corrector.

In \cite{mazumdar2021support}, a symbol energy based false alarm corrector is designed where the false detected users are recognized through detecting the energy of users' decoded symbols. Nevertheless, such a false alarm corrector is susceptible to noise. 
However, in our case, a different false alarm corrector based on multiple zero symbol detection \cite{zhu2010exploiting} is developed. The key idea is that if a user is active, the detected zero symbol number in its decoded packet $\hat{\mathbf{x}}_{u}$ should be small, otherwise, the  detected zero symbol number should be large. Let $\tau_{zs}\geq 1 \in \mathbf{Z}_{+}$ denotes the threshold of the  detected zero symbol number in any one data packet of users. For user $u, u\in \hat{\mathcal{U}}_{ac}^{I}$, the proposed false alarm corrector can be summarized as
\begin{equation}
\hat{a}_{u}=\left\{\begin{array}{l}
0,\quad  K-||\hat{\mathbf{x}}_{u}||_{0}>=\tau_{zs}\\
1, \quad {\rm otherwise}
\end{array}\right.
\end{equation}
where $\vert\vert \cdot \vert\vert_{0}$ denotes the $l_{0}$-norm. When $\tau_{zs}=1$, our false alarm corrector is the same as that in \cite{zhu2010exploiting}. In practical applications, a smaller $\tau_{zs}$ would result in more missing detection, while a bigger $\tau_{zs}$ would result in more false alarms which may exceed the tolerance of the  false alarm corrector. In this paper, to balance these two performances, the value of $\tau_{zs}$ is chosen as $\lceil \frac{K}{3}  \rceil$ where $\lceil \cdot \rceil$ denotes rounding up to the nearest integer.
The pseudo-code of our proposal is given in Algorithm 1.
\begin{algorithm}[htbp]
	\caption{Data aided active user detection}
\begin{algorithmic}[1] \footnotesize
\Require $\mathbf{y}[:,k], k\in[1,K]$, $\mathbf{y}_{s}$
\Ensure $\hat{\mathcal{U}}_{ac}^{II}$, $\hat{\mathbf{x}}_{u}[k], u\in \hat{\mathcal{U}}_{ac}^{II}, k\in [1,K]$
\State Estimate $\hat{U}_{ac}^{I}$ by MPA detector in (8) - (11) or TL-MPA in (12) - (16); \quad\quad\quad\quad\quad\quad\quad\quad //step 1
\State Construct factor graph $\mathcal{G}(\hat{\mathcal{U}}_{ac}^{I})$;  
\For {$k=1:K$}
	\State  $\hat{\mathbf{x}}_{u}[k] = {\rm MPA}( \mathbf{y}[:,k], \mathcal{G}(\hat{\mathcal{U}}_{ac}^{I}) ), u\in \hat{\mathcal{U}}_{ac}^{I}$;
\EndFor
\For {$\forall u\in \hat{\mathcal{U}}_{ac}^{I}$} \quad\quad\quad //step 2
	\If { $K - \vert\vert \hat{\mathbf{x}}_{u} \vert\vert_{0} \geq \tau_{zs}$}
		$\hat{a}_{u}$ = 0;
		$\hat{\mathcal{U}}_{ac}^{I}=\hat{\mathcal{U}}_{ac}^{I}-\{u\}$;
	\EndIf
\EndFor
\State $\hat{\mathcal{U}}_{ac}^{II}=\hat{\mathcal{U}}_{ac}^{I}$; 
\State Construct factor graph $\mathcal{G}(\hat{\mathcal{U}}_{ac}^{II})$;
\For {$k=1:K$}
\State  $\hat{\mathbf{x}}_{u}[k] = {\rm MPA}( \mathbf{y}_{k}, \mathcal{G}(\hat{\mathcal{U}}_{ac}^{II}) ), u\in \hat{\mathcal{U}}_{ac}^{II}$;
\EndFor
\end{algorithmic}
\end{algorithm}

\section{Complexity analysis}
\begin{figure}[htbp]
	\centering
	\includegraphics[scale=0.2]{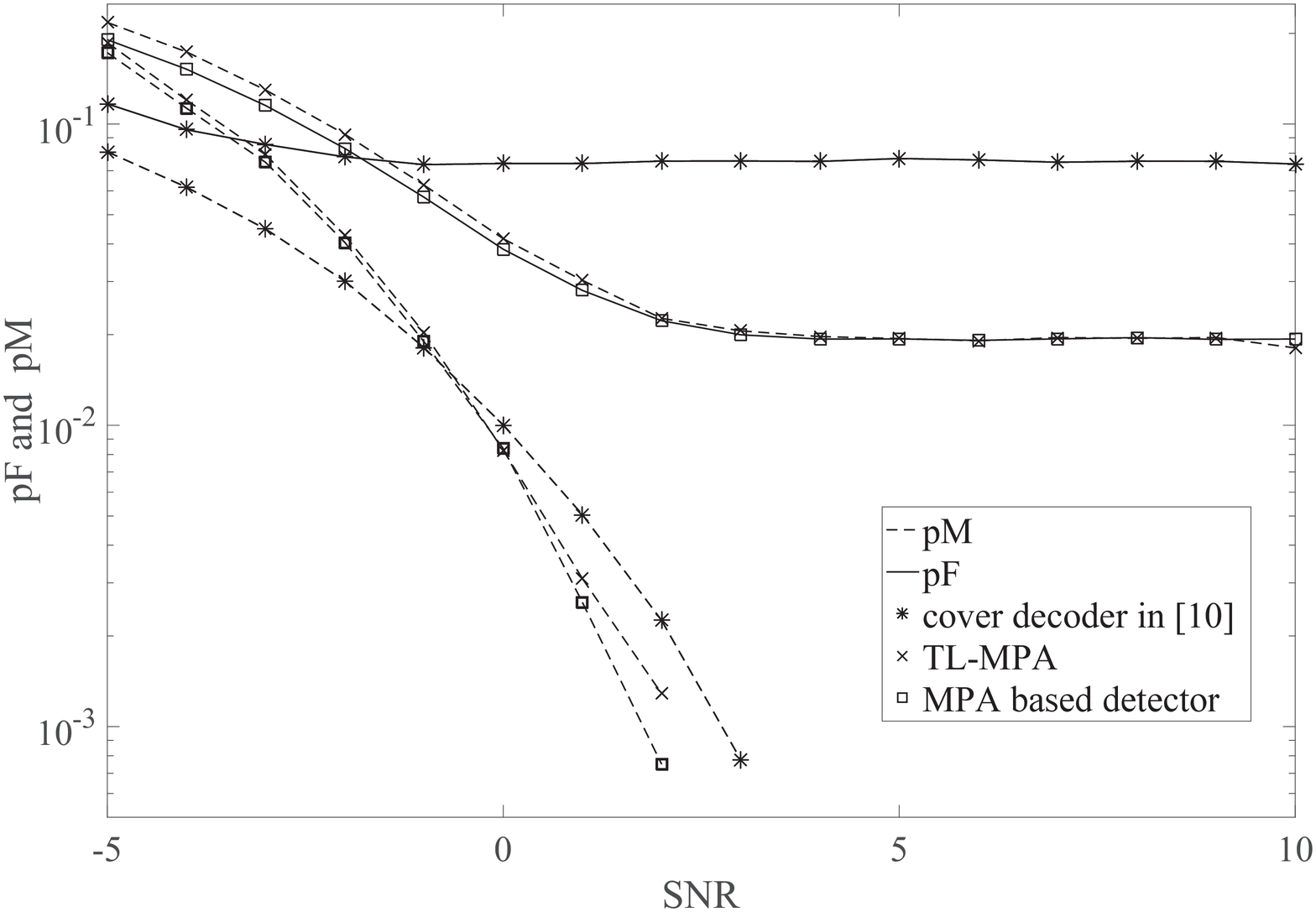}
	\caption{ $\hat{\mathcal{U}}_{ac}^{I}$ estimation comparison between the proposed MPA based detector, TL-MPA based detector and cover decoder in \cite{inan2019group}, where $\lambda=0.1$.}
	\label{2}
\end{figure}

Instead of the perfect factor graph $\mathcal{G}(\mathcal{U}_{ac})$ \cite{oyerinde2019compressive,wang2016dynamic}, executing MPA over the factor graph $\mathcal{G}(\hat{\mathcal{U}}_{ac}^{I})$ will not increase the complexity order of MPA in the data decoding part, because the false alarms in $\hat{\mathcal{U}}_{ac}^{I}$ are small. This fact is revealed later in Fig. \ref{2}. 
Hence, we mainly compare the complexity of the AUD part in this section.

The complexity order of OMP and AMP are analyzed in Table I in our previous work \cite{yang2021cross}. The complexity of DCS is approximately in the same order as OMP.   
Dominated by (8), the complexity of the MPA based  detector $\mathcal{C}_{\rm MPA}$ is in the order of $\mathcal{O}(L_{s}2^{w_{r}})$. The complexity of TL-MPA is dominated by the degree distribution of check nodes in $\mathcal{G}(\hat{\mathcal{U}}_{ac}^{I})$  which can be well approximated by
\begin{equation}
\mathcal{C}_{\rm TL-MPA} \approx \mathcal{O}(L_{s}( p_{w_{1}} \binom{w_{1}}{w_{r}}  + p_{w_{2} }\binom{w_{2}}{w_{r}}  )  ),
\end{equation}
where $p_{w_{1}}=1-[\lambda w_{r}-\lfloor \lambda w_{r} \rfloor]$, $w_{1}=\lfloor \lambda w_{r} \rfloor$, $p_{w_{2}}= 1- p_{w_{1}}$, and $w_{2}=w_{1}+1$. 
Finally, the complexity orders of other algorithms are listed in Table I
\begin{table}[htbp]
	\caption{Complexity comparison}
	\begin{center}
		\begin{tabular}{ c|c } 
			\hline
		 Algorithm  & Complexity order   \\
		 \hline
	DCS and	OMP in \cite{oyerinde2019compressive,wang2016dynamic}      &    $\mathcal{O}(N_{a}L_{s}N+N_{a}^{3}+N_{a}L_{s})$            \\
		\hline
		AMP in \cite{yang2021cross,8264818}       &    $\mathcal{O}(L_{s}N)$ \\
			\hline
			 MPA based detector & $\mathcal{O}(L_{s}2^{w_{r}})$    \\
			\hline
		TL-MPA based detector  & $\mathcal{O}(L_{s}( p_{w_{1}} \binom{w_{1}}{w_{r}}  + p_{w_{2} }\binom{w_{2}}{w_{r}}  )  )$  \\
		\hline
		\end{tabular}
	\end{center}
\end{table}

\section{Simulation Results and Discussion}
In this section, the AUD performance and payload data decoding performance are simulated. 
To evaluate the AUD performance, the probability of miss detection (pM) and the probability of false detection (pF) are adopted
\cite{yang2021cross,8264818}.
To measure the data decoding performance,
the symbol error rate (SER) is adopted \cite{jiang2020joint}.
The system configuration is given in TABLE II. 
\begin{table}[htbp]
	\caption{System Configuration}
	\begin{center}
		\begin{tabular}{ c|c} 
			\hline
			Potential user number $N$  & 80   \\
			\hline
			User sparsity $\lambda$ & 0.1, 0.3 \\
			\hline
			Sub-carrier number $L_{s}$     &   39 \\
			\hline
			The value of $w_{c}$ &2 \\
			\hline
			The value of $w_{r}$ &4 \\
			\hline
			The packet length $K$ & 10 \\
			\hline
			Constellation alphabet size $M$ & 2\\
			\hline
		\end{tabular}
	\end{center}
\end{table}

Firstly, the quality of super-set $\hat{\mathcal{U}}_{ac}^{I}$ estimated by the proposed MPA based initial estimator and TL-MPA based initial estimator are evaluated.
The pF performances of MPA based detector and TL-MPA based detector outperform  cover decoder in \cite{inan2019group} significantly when $\rm SNR> 0$dB. The main reason is that more specific traffic load information $\hat{\mathbf{Y}}_{l}$ is utilized by both the MPA based detector and the TL-MPA based detector.
Moreover, the performance of TL-MPA is only slightly worse than MPA detector when ${\rm SNR<3}$ dB.
This phenomenon verifies the efficiency of the proposed TL-MPA based detector scheme.

\begin{figure}[htbp]
	\centering
	\includegraphics[scale=0.2]{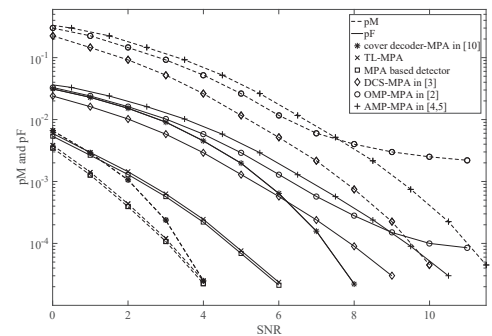}
	\caption{AUD performance comparison between the proposed data aided AUD scheme and its counterparts in \cite{wang2016dynamic,oyerinde2019compressive,yang2021cross,8264818,inan2019group} where $\lambda = 0.1$.}
	\label{3}
\end{figure}
The AUD performance of our proposed method with $\lambda=0.1$ is shown in Fig. \ref{3}. OMP-MPA indicates that the AUD is performed by OMP algorithm and data decoding is performed by MPA represented in (17). In the same spirit, we have AMP-MPA, DCS-MPA. Cover decoder-MPA denotes that estimating $\hat{\mathcal{U}}_{ac}^{I}$ by cover decoder based on $\hat{\mathbf{Y}}_{t}$ in (6) and data decoding is performed by MPA. 
The pF performance of cover decoder-MPA is almost the same as that of DCS-MPA which has the best AUD performance in CS based counterparts. 
The pM performance of the cover decoder-MPA is significantly better than that of the DCS-MPA.
Owing to the higher quality of super-set estimation $\hat{\mathcal{U}}_{ac}^{I}$, the pF performances of the proposed MPA based detector and the TL-MPA based detector outperform cover decoder-MPA significantly.

\begin{figure}[htbp]
	\centering
	\includegraphics[scale=0.2]{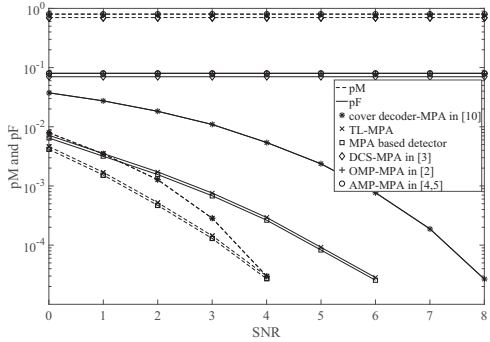}
	\caption{AUD performance comparison between proposed data aided AUD scheme and its counterparts in \cite{wang2016dynamic,oyerinde2019compressive,yang2021cross,8264818,inan2019group} where $\lambda = 0.3$.}
	\label{4}
\end{figure}
The AUD performance of our proposed method in a relatively higher user sparsity region, i.e. $\lambda=0.3$, is shown in Fig. \ref{4}. The performance of CS-MPA detectors in \cite{wang2016dynamic,oyerinde2019compressive,yang2021cross} is poor in this sparsity due to the sparsity limitation in CS theory, whereas our proposed method works well. It implies that more active users can be supported by our method.

\begin{figure}[htbp]
	\centering
	\includegraphics[scale=0.2]{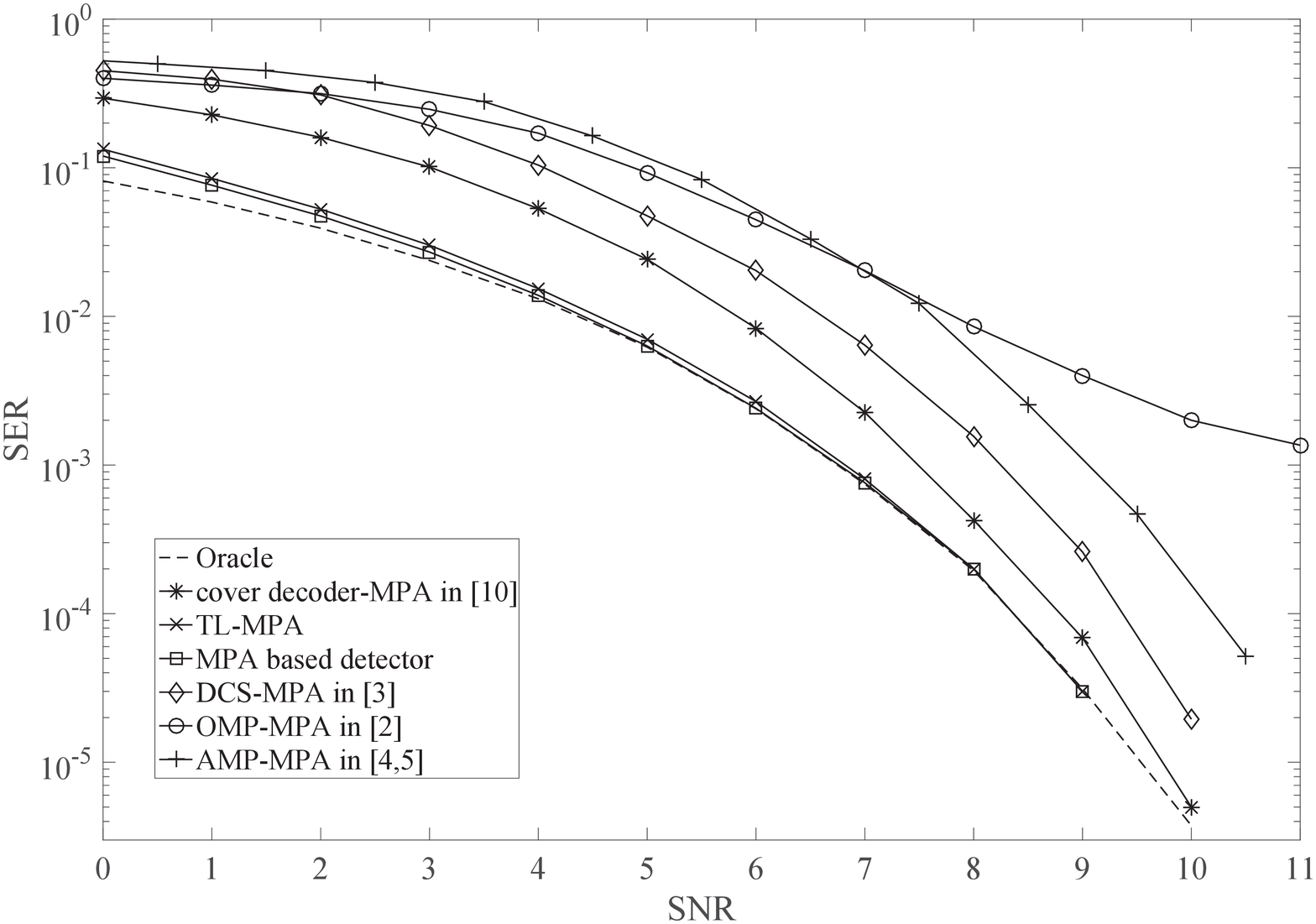}
	\caption{SER performance comparison between proposed data aided AUD scheme and its counterparts in \cite{wang2016dynamic,oyerinde2019compressive,yang2021cross,8264818,inan2019group} where $\lambda = 0.1$.}
	\label{5}
\end{figure}

The SER performance of our proposed method with $\lambda=0.1$ is shown in Fig. \ref{5}. 
The oracle performance means the SER performance of MPA with perfect AUD.
Owing to the superior AUD performance represented in Fig. \ref{3}, our proposed method achieves the best SER performance. 

\begin{figure}[htbp]
	\centering
	\includegraphics[scale=0.2]{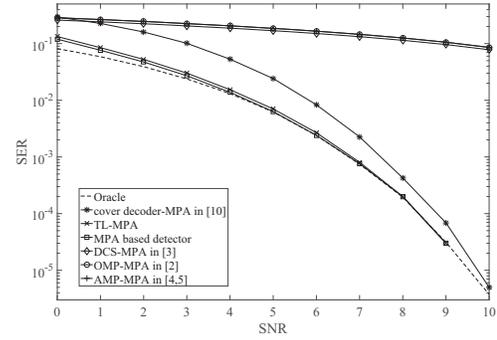}
	\caption{SER performance comparison between proposed data aided AUD scheme and its counterparts in \cite{wang2016dynamic,oyerinde2019compressive,yang2021cross,8264818,inan2019group} where $\lambda = 0.3$.}
	\label{6}
\end{figure}
The SER performance of our proposed method with $\lambda=0.3$ is shown in Fig. \ref{6}. Similar to Fig. \ref{4}, it confirms that many more active users can be supported by our method.

\section{Conclusion}
In this paper, we transfer the AUD problem as a super-set estimation problem  based on the observation that the false detected users could be possibly corrected with the aid of decoded data symbols.
Then, a two-step data-aided AUD scheme with false alarm correction is proposed. 
To estimate an initial active user set in step 1, the embedded LDS based user preamble pool is constructed and two MPA based initial estimators are developed to realize the detection.
In addition, a false alarm corrector is integrated into the data decoding stage to recognize the remaining false detected inactive users in the initial active user set.
Simulation results verify the efficiency and superior performance of our proposed methods.

\bibliographystyle{IEEEtran}
\bibliography{IEEEabrv,ForIEEEBib}

\begin{thebibliography}{10}
\providecommand{\url}[1]{#1}
\csname url@samestyle\endcsname
\providecommand{\newblock}{\relax}
\providecommand{\bibinfo}[2]{#2}
\providecommand{\BIBentrySTDinterwordspacing}{\spaceskip=0pt\relax}
\providecommand{\BIBentryALTinterwordstretchfactor}{4}
\providecommand{\BIBentryALTinterwordspacing}{\spaceskip=\fontdimen2\font plus
\BIBentryALTinterwordstretchfactor\fontdimen3\font minus
  \fontdimen4\font\relax}
\providecommand{\BIBforeignlanguage}[2]{{%
\expandafter\ifx\csname l@#1\endcsname\relax
\typeout{** WARNING: IEEEtran.bst: No hyphenation pattern has been}%
\typeout{** loaded for the language `#1'. Using the pattern for}%
\typeout{** the default language instead.}%
\else
\language=\csname l@#1\endcsname
\fi
#2}}
\providecommand{\BIBdecl}{\relax}
\BIBdecl

\bibitem{shahab2020grant}
M.~B. Shahab, R.~Abbas, M.~Shirvanimoghaddam, and S.~J. Johnson, ``Grant-free
  non-orthogonal multiple access for iot: A survey,'' \emph{IEEE Communications
  Surveys \& Tutorials}, vol.~22, no.~3, pp. 1805--1838, 2020.

\bibitem{oyerinde2019compressive}
O.~O. Oyerinde, ``Compressive sensing algorithms for multiuser detection in
  uplink grant free {NOMA} systems,'' in \emph{2019 IEEE 89th Vehicular
  Technology Conference (VTC2019-Spring)}.\hskip 1em plus 0.5em minus
  0.4em\relax IEEE, 2019, pp. 1--6.

\bibitem{wang2016dynamic}
B.~Wang, L.~Dai, Y.~Zhang, T.~Mir, and J.~Li, ``Dynamic compressive
  sensing-based multi-user detection for uplink grant-free {NOMA},'' \emph{IEEE
  Communications Letters}, vol.~20, no.~11, pp. 2320--2323, 2016.

\bibitem{yang2021cross}
L.~Yang, P.~Fan, L.~Li, Z.~Ding, and L.~Hao, ``Cross validation aided
  approximated message passing algorithm for user identification in {mMTC},''
  \emph{IEEE Communications Letters}, vol.~25, no.~6, pp. 2077--2081, 2021.

\bibitem{8264818}
Z.~Chen, F.~Sohrabi, and W.~Yu, ``Sparse activity detection for massive
  connectivity,'' \emph{IEEE Transactions on Signal Processing}, vol.~66,
  no.~7, pp. 1890--1904, 2018.

\bibitem{zhu2010exploiting}
H.~Zhu and G.~B. Giannakis, ``Exploiting sparse user activity in multiuser
  detection,'' \emph{IEEE Transactions on Communications}, vol.~59, no.~2, pp.
  454--465, 2010.

\bibitem{bian2021supporting}
X.~Bian, Y.~Mao, and J.~Zhang, ``Supporting more active users for massive
  access via data-assisted activity detection,'' in \emph{ICC 2021-IEEE
  International Conference on Communications}.\hskip 1em plus 0.5em minus
  0.4em\relax IEEE, 2021, pp. 1--6.

\bibitem{van2009multiple}
J.~Van De~Beek and B.~M. Popovic, ``Multiple access with low-density
  signatures,'' in \emph{GLOBECOM 2009-2009 IEEE Global Telecommunications
  Conference}.\hskip 1em plus 0.5em minus 0.4em\relax IEEE, 2009, pp. 1--6.

\bibitem{mazumdar2021support}
A.~Mazumdar and S.~Pal, ``Support recovery in universal one-bit compressed
  sensing,'' \emph{arXiv preprint arXiv:2107.09091}, 2021.

\bibitem{inan2019group}
H.~A. Inan, S.~Ahn, P.~Kairouz, and A.~Ozgur, ``A group testing approach to
  random access for short-packet communication,'' in \emph{2019 IEEE
  International Symposium on Information Theory (ISIT)}.\hskip 1em plus 0.5em
  minus 0.4em\relax IEEE, 2019, pp. 96--100.

\bibitem{jang2016early}
H.~S. Jang, S.~M. Kim, H.-S. Park, and D.~K. Sung, ``An early preamble
  collision detection scheme based on tagged preambles for cellular m2m random
  access,'' \emph{IEEE Transactions on Vehicular Technology}, vol.~66, no.~7,
  pp. 5974--5984, 2016.

\bibitem{hoshyar2008novel}
R.~Hoshyar, F.~P. Wathan, and R.~Tafazolli, ``Novel low-density signature for
  synchronous {CDMA} systems over {AWGN} channel,'' \emph{IEEE Transactions on
  Signal Processing}, vol.~56, no.~4, pp. 1616--1626, 2008.

\bibitem{jiang2020joint}
S.~Jiang, X.~Yuan, X.~Wang, C.~Xu, and W.~Yu, ``Joint user identification,
  channel estimation, and signal detection for grant-free {NOMA},'' \emph{IEEE
  Transactions on Wireless Communications}, vol.~19, no.~10, pp. 6960--6976,
  2020.

\end{thebibliography}

\end{document}